\documentstyle[aas2pp4,psfig]{article}
\slugcomment{Submitted 19 April 2001; Accepted 19 July 2001}

\lefthead{Smale}
\righthead{Burst in X2127$+$119}

\def\source{X2127$+$119}

\def\approxlt{\mathrel{\hbox{\rlap{\lower.55ex \hbox {$\sim$}}
        \kern-.3em \raise.4ex \hbox{$<$}}}}
\def\approxgt{\mathrel{\hbox{\rlap{\lower.55ex \hbox {$\sim$}}
        \kern-.3em \raise.4ex \hbox{$>$}}}}

\begin{document}

\title{A Second Intense Burst with Photospheric Radius Expansion from
\source\ in M15}

\author{Alan P. Smale\altaffilmark{1}}
\affil{Laboratory for High Energy Astrophysics,
Code 662, NASA/Goddard Space Flight Center, Greenbelt, MD~20771}

\altaffiltext{1}{Also Universities Space Research Association.} 

\begin{abstract}

In 2000 September we observed a bright X-ray burst from \source\ with
the {\sl Rossi X-ray Timing Explorer}. This burst has a multi-peaked
profile, a peak luminosity of $\sim$6.5$\times$10$^{38}$~erg~s$^{-1}$,
a total integrated energy of $\sim$2$\times$10$^{40}$~ergs, and
significant photospheric radius expansion to a maximum extent of
$R$=118$\pm$5~km. From the luminosity-temperature relation during the
expansion phase we derive estimates for the gravitational redshift at
the neutron star surface, the corrected Eddington luminosity, and the
neutron star mass.  We detect no slow ($\sim$30s) radial oscillations
or fast (100--1200~Hz) coherent oscillations or QPO during the burst.

The 2000 September event is only the second burst ever observed from
this globular cluster binary (in M15 = NGC~7078), and it shares many
characteristics with the event detected by Ginga in 1988 October
(Dotani et al.\ 1990; van Paradijs et al.\ 1990), the key difference
probably being the total amount of material consumed in the
thermonuclear flash.

\end{abstract}


\section{Introduction}

The X-ray source \source\ is a bright low-mass binary located within
2'' of the core of the globular cluster M15 (=NGC~7078). It is
identified with the V$\sim$15 optical star AC211 (Auri\`ere et al.\
1984), which is far and away the brightest counterpart of any globular
cluster X-ray source. When folded on the 17.11-hr optical period
(Ilovaisky et al.\ 1993; Homer \& Charles 1998), the light curve
of \source\ shows a
broad symmetrical $\sim$5-hr eclipse at phase $\phi$=0, and
the largest orbital modulation seen from any X-ray binary (1.8 mag in U), but
with extremely erratic and complex X-ray and optical behavior between
phases $\phi$=0.2--0.7. Combining this variability with its low
$L_x/L_{opt}$ ratio of $\sim$20, the natural conclusion is that
\source\ is an accretion disk corona (ADC) source observed at a
sufficiently high inclination that the bulk of the direct X-ray
emission is obscured by azimuthal disk structure. However, the
detection of a very luminous X-ray burst from the source with the
Ginga satellite (Dotani et al.\ 1990; van Paradijs et al.\ 1990;
hereafter ``the 1988 burst'') not only revealed that the compact
object must be a neutron star, but also that it must be viewed
directly with no intervening obscuration.

The 1988 burst from \source\ had a long duration (in excess of 150s),
a peak luminosity of 5.1$\times$10$^{38}$ erg~s$^{-1}$, and a total
energy release of 6.3$\times$10$^{40}$ ergs (correcting the distance
to 10.4~kpc; see below), placing it amongst the brightest of the
observed Type I X-ray bursts. A bright precursor was seen 6 seconds
prior to the main event, and spectral fitting revealed strong
photospheric radius expansion during the burst, followed by
contraction and a cooling phase. In the first 30s of this contraction,
slow radius oscillations occurred (Dotani et al.\ 1990; van Paradijs
et al.\ 1990).

Per unit mass, 
low mass X-ray binaries (LMXBs) are $\sim$100 times more likely to be
found in globular clusters than in the Galaxy as a whole, and they are
almost certainly formed by the tidal capture of compact objects, an
efficient process within globular clusters because of the high local
number density of stars (Fabian, Pringle, \& Rees 1975; Hut et al.\
1992). The bursts observed from the 12 known
globular cluster LMXBs $>10^{36}$ erg~s$^{-1}$ have a special
significance because we have relatively accurate information about (a)
their distance, and (b) the elemental abundances of the regions in
which they were formed. The M15 cluster has a low mean
metallicity of $\sim$0.01 Solar (Sneden et al.\ 1991; Geisler et
al.\ 1992; Harris 1996), although the local metallicity of the accreted
material in the \source\ system may, of course, have been altered from
this by evolutionary effects.

In this Paper, we report the observation and analysis of a second
burst from \source, 12 years after the first. The burst reported here
shares many of the characteristics of the earlier event, being highly
energetic with radius expansion, a high peak luminosity, and a
multi-peaked profile; however, its main energy release is of shorter
duration, leading to a lower total energy, and no precursor is
observed.

\section{Observations and Results}

We observed \source\ with {\sl RXTE} (Bradt, Rothschild, \& Swank,
1993) on 2000 August 24 04:36--10:31~UT, and then for a series of
pointings spanning 2000 September 17 12:02~UT -- September 23
02:36~UT, for an on-source total good time of 150~ksec.  The data
presented here were obtained using the ``Standard 2'' mode with 16s
time resolution, and an event mode with 16$\mu$sec time resolution and
64 channels covering the full energy range of the PCA.  The PCA
consists of five Xe Proportional Counter Units (PCUs) numbered PCUs 0
through 4, with a combined total effective area of about 6500 cm$^2$
(Jahoda et al.\ 1996). PCUs 0, 2, and 3 were reliably on throughout the
observation, and the data shown here are from these detectors. Data
were reduced using the {\sl RXTE} standard analysis software included
in FTOOLS 5.0.1, and we used XSPEC 11.0 for our spectral
analysis. Where necessary a small instrument-related deadtime
correction (6\%) has been applied to the derived PCA fluxes. Errors on
spectral fit parameters are quoted to 90\% confidence. 

\begin{figure}[htb]
\figurenum{1}
\begin{center}
\begin{tabular}{c}
\psfig{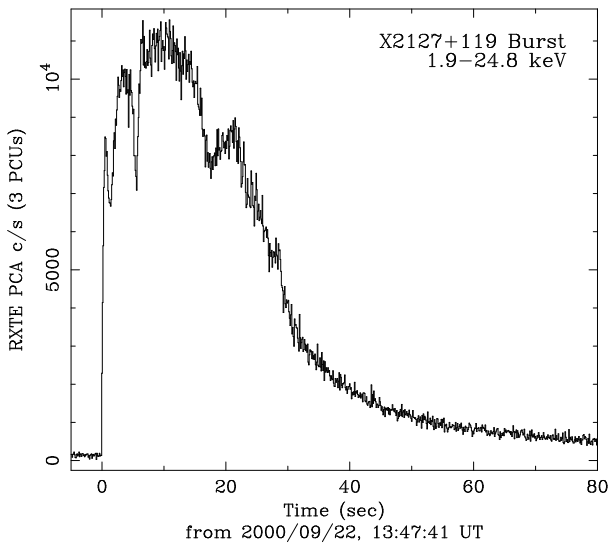}
\end{tabular}
\caption{The intensity profile of the bright Type I burst observed from
\source\ in 2000 September, plotted with a time resolution of 0.1s.
}
\end{center}
\end{figure}

The burst was observed during the September sequence, at 2000
September 22 13:47:41~UT, a time corresponding to phase
$\phi$=0.79$\pm$0.03 in the binary cycle of \source, well away from
the partial eclipse. In Figure~1 we show the intensity profile of the
burst in the 2--25~keV range. The profile has a complex structure,
with an initial fast rise to a first emission peak in $<$0.5s,
followed by three further peaks in the overall counting rate separated
from the first by 3.75s, 11.0s, and 21.5s (all such times in this
paper are quoted to $\pm$0.25s accuracy unless otherwise stated).
The burst took the source
from its persistent emission rate of 90 c/s to a maximum of 11,000
c/s, after which it decayed with a {\sl 1/e}-folding time of 30s. The
effects of the burst are still visible in the X-ray spectra 300s
after its initial rise. A full timing analysis was performed on the
burst data: no slow ($\sim$30s) radius oscillations or fast
(100~Hz--1200~Hz) coherent oscillations or QPO were detected.

The energy dependence of the burst profile is illustrated in Figure~2.
In the 1.9--3.5~keV band the initial energy release appears gradual,
extending over 6s. At the highest energies the burst appears as
a bright 1s pulse, followed by a more gradual 20s rise and an
almost-symmetrical decay. Such energy-dependent double peaking in Type
I bursts has long been recognized as the signature of photospheric
radius expansion in bursts reaching the Eddington luminosity (Hoffman,
Cominsky, \& Lewin 1980; Lewin, van Paradijs, \& Taam 1993).

\begin{figure}[htb]
\figurenum{2}
\begin{center}
\begin{tabular}{c}
\psfig{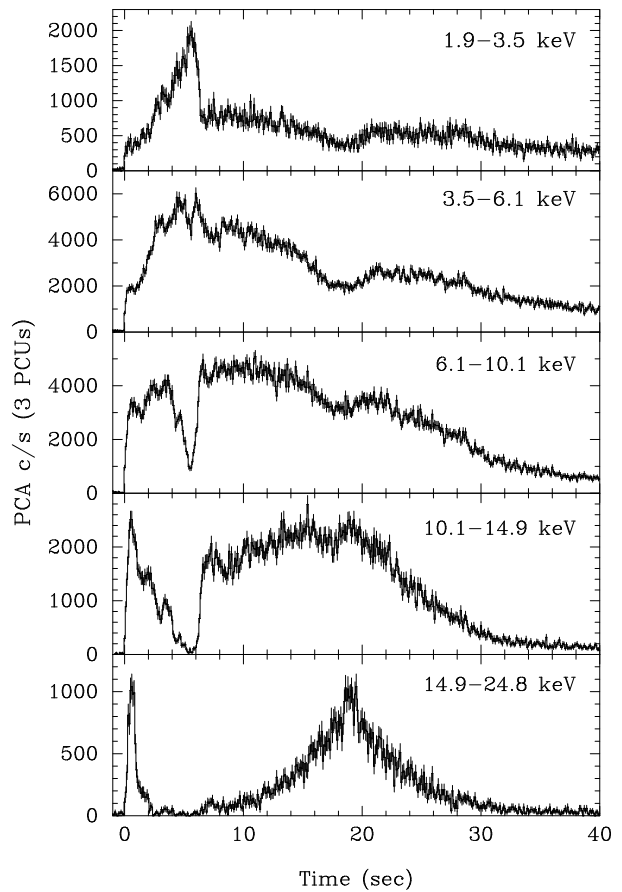}
\end{tabular}
\caption{The burst light curve, broken down into several bands to
show the energy dependence of the profile.
}
\end{center}
\end{figure}

To study the physical morphology of the burst in more detail, we
extracted a series of spectral slices through the event with a time
resolution of 0.25s, and performed spectral fits to them over the
2.5--15~keV energy range.  Our first pass ultimately proved to be the
most satisfactory: we fit a simple blackbody function to the burst
spectra, using as background a 300s section of persistent-emission
data from immediately prior to the burst. Experimentation showed that
the hydrogen column density was poorly constrained by the data,
leading us to fix it at a nominal value of $N_H=7\times$10$^{21}$
cm$^{-2}$, consistent with previous observations
(e.g. Sidoli et al.\ 2000). These fits produced acceptable values of
$\chi^2$ throughout the burst.  No improvements were gained using more
sophisticated fitting models that include Comptonization of the
blackbody component or modification of the blackbody by electron
scattering (Nishimura, Mitsuda, \& Itoh 1986; Nakamura et al. 1989),
or a contribution from disk reflection of the burst emission (Day \&
Done 1991). We saw no indications of transient narrow absorption or
emission features in the spectra.

In Figure~3, we plot the variation of the bolometric luminosity, the
blackbody temperature $kT_{bb}$, and the derived apparent blackbody
radius $R_{bb}$ through the burst as a function of time. (At the
temperatures and densities found on accreting neutron stars, the
Compton scattering opacity dominates over the absorption opacity,
reducing the emissivity below that expected for a true
blackbody. Thus, this method measures the color temperature rather than the
effective temperature $kT_{eff}$. We will return to this point in the
Discussion section.) In the lowest panel we see clear evidence for
photospheric radius expansion; the apparent radius increases in 5.6s
to a maximum envelope radius of 
$R_{bb}$=118$\pm$5~km, followed by an
even more rapid collapse to 
$R_{bb}$=29.0$\pm$1.5~km over the next
1.0s. After this point the apparent radius contracts at a more steady
rate to a minimum value of 
8.6$\pm$1.0~km, 20s after the initial burst rise.  
As is common in radius expansion bursts, the measured blackbody
temperature mirrors the behavior of the derived radius. After an
initial temperature peak during the first 1.0s of the burst, the
temperature drops sharply during the photospheric expansion
phase. During the subsequent contraction, the temperature climbs back
to its peak value of 
$kT_{bb}$=2.70$\pm$0.13~keV, and then cools
exponentially as the blackbody radius stabilizes. In calculating the
luminosity we assume that the radiation is isotropic and that the
source is located at a distance of $d$=10.4~kpc (Durrel \& Harris
1993; Harris 1996). The luminosity peaks at a value of
6.5$\times$10$^{38}$~erg~s$^{-1}$, and over the whole 300s interval
where the effects of the burst are detectable, the total integrated
burst energy is 1.85$\times$10$^{40}$~ergs.

\begin{figure}[htb]
\figurenum{3}
\begin{center}
\begin{tabular}{c}
\psfig{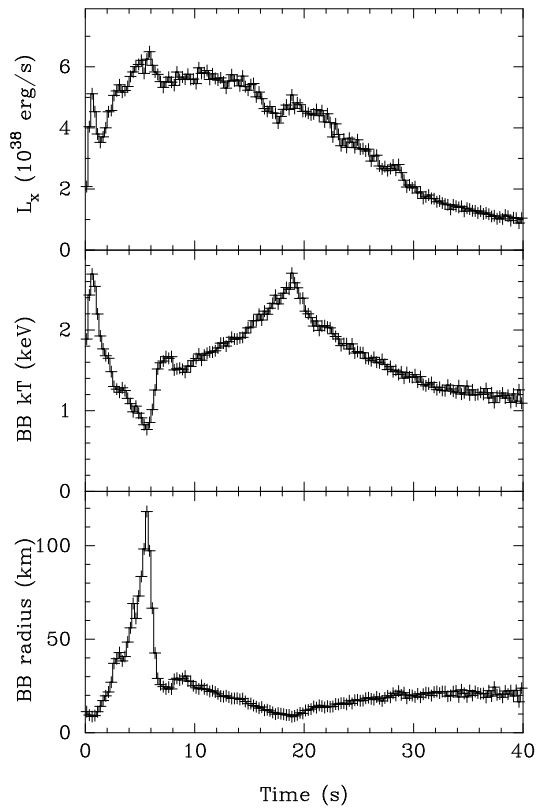}
\end{tabular}
\caption{The variation of the bolometric luminosity (for a distance
of 10.4 kpc), blackbody color temperature $kT_{bb}$, and apparent
blackbody radius $R$, derived from fits to spectra accumulated through
the burst with 0.25s resolution.
}
\end{center}
\end{figure}

While it is common practice in burst spectral studies to subtract off
the continuum emission prior to fitting, it is well-known that this is
a source of potential error (e.g. Lewin et al.\ 1993).
If there are two components to the persistent emission, and one
component originates on the neutron star or in the boundary layer
around it, this component of the emission may either increase or
disappear during the burst itself, meaning that the ``continuum
background'' is either underestimated or overestimated. This can lead
to an erroneous decrease in the apparent blackbody radius particularly
towards the
end of the burst, where the flux level is low.  In addition, for
bursts that approach or exceed the Eddington limit, the possibility
exists that the global accretion onto the neutron star itself may be
suppressed, leaving the burst itself as the {\sl only} source of X-ray
emission for the brief duration of the burst.

We investigated both of these possibilities to determine whether they
affect our results.  The persistent continuum emission prior to the
burst (and, in fact, through the whole RXTE observation) can be well
fit using either a straightforward cut-off power law with index
$\Gamma$=1.63$\pm$0.16, $E_{cut}$=8.2$\pm$1.8~keV, $\chi^2$=26/36, or
with a combination of a powerlaw plus blackbody, with
$\Gamma$=2.42$\pm$0.11, $kT_{bb}$=1.43$\pm$0.16~keV, $\chi^2$=20/35, with
the blackbody providing $\sim$20\% of the total flux. In the first
formulation, there is no contribution from the region on or around the
neutron star; adopting the second, we repeated our series of fits
through the burst, holding the powerlaw component constant and
allowing the continuum$+$burst blackbody component to vary. We found
that the continuum blackbody contribution was a sufficiently small
addition that it had a negligible effect on Figure~3 and on the
physical numbers quoted below.  (The broadband 0.1--100 keV spectrum
of \source\ was found to be well-described by a partially-covered
powerlaw plus a disk blackbody (Sidoli et al.\ 2000); our spectroscopy
results are consistent with this model, though the PCA lacks the
sensitivity to the lowest energies provided by the SAX LECS instrument
that would enable us to perform a detailed study of partial covering
over the limited intensity variations found in these data.)

We next repeated our series of burst fits assuming a cessation of
persistent emission during the burst, and fitting a blackbody to the
overall emission with only the relevant instrumental PCA background
subtracted. This produced poor fits to the individual spectra, with
large systematic errors in the residuals, indicating that the
assumption was invalid and that persistent emission does indeed
continue unabated during the burst.

\section{Discussion}

A comparison of the 1988 and 2000 bursts from \source\ shows that
there are strong similarities between the two events. Both show
multiple peaks in their X-ray light curves in count- and flux- space,
and both show indisputable evidence for radius expansion and a
qualitatively similar evolution of flux and blackbody temperature and
radius. The bursts have peak luminosities of
$L_{max}$=5.3$\times$10$^{38}$ erg~s$^{-1}$ and 6.5$\times$10$^{38}$
erg~s$^{-1}$ respectively; peak blackbody temperatures of
$kT_{bb}$=2.87$\pm$0.03 keV and $kT_{bb}$=2.70$\pm$0.13 keV; and developed
from comparable levels of continuum emission
($F_{pers}$=5.0$\times$10$^{-10}$~erg~cm$^{-2}$~s$^{-1}$ and
\break
$F_{pers}$=4.2$\times$10$^{-10}$~erg~cm$^{-2}$~s$^{-1}$ over the
1--28~keV range) (Dotani et al.\ 1990, van Paradijs et al.\ 1990, and
above, assuming $d$=10.4~kpc).

The bursts differ in the maximum extent of their photospheric
expansion -- $R_{bb}\sim$1000 km and $R_{bb}\sim$120 km respectively --
and in their timescales.  The 1988 burst reached its maximum radius
expansion after $\sim$6s, the end of its sharp radius contraction
after $\sim$18s, and final photospheric touchdown (as defined by the
simultaneous peak in $kT_{bb}$ and trough in $R_{bb}$) after
$\sim$90s.  For the 2000 burst these times are respectively 5.6s,
6.6s, and $\sim$19s.  The e-folding time for the decay of the 1988
burst is hard to determine from van Paradijs et al.\ (1990), but they
note that the burst intensity had decreased to 20\% by 160s after
burst onset. For the 2000 burst, the 20\% mark is reached after
38s. Largely due to this difference in timescales, the total
bolometric energy released by the bursts differs by a factor of 4:
6.3$\times$10$^{40}$ ergs for the 1988 event, as compared to
1.85$\times$10$^{40}$ ergs for the 2000 event. Although a little
weaker than the first burst, the burst described here still ranks as
an unusually energetic event.

From the fluxes derived earlier we calculate the burst characteristics
$\gamma$ (=$F_{pers}$/$F_{max}$) and $\tau$ (=$E_b$/$F_{max}$, where
$E_b$ is the total burst fluence or integrated net flux across the burst) to
be 0.0125 and 125s respectively for the 1988 burst, and 0.0084 and 28.5s 
for the
burst presented above.  In their summary of the properties of
Eddington-limited bursts from a large number of sources, van Paradijs,
Penninx \& Lewin (1988) find a strong anti-correlation between
log~$\tau$ and log~$\gamma$ for Type I bursts showing photospheric
radius expansion. The \source\ bursts lie on the line of this
anti-correlation, but at the extreme end of the ($\gamma, \tau$)
distribution, indicating the intensity and
long duration of the bursts from \source\ relative to those from other
sources.

\begin{figure}[htb]
\figurenum{4}
\begin{center}
\begin{tabular}{c}
\psfig{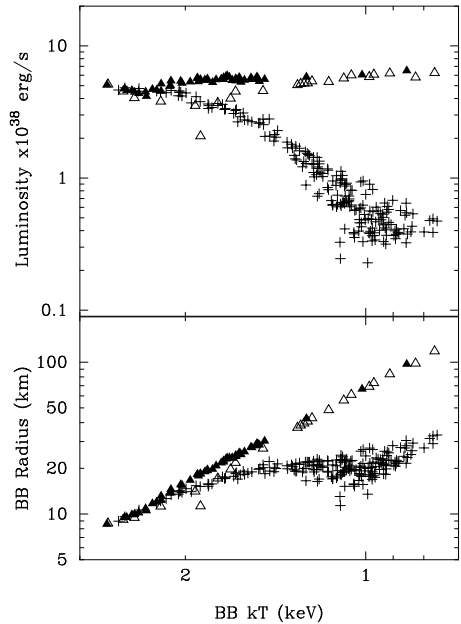}
\end{tabular}
\caption{The data from Figure 3, extended out to 100s after the burst
rise, and replotted to show the (a) luminosity-temperature diagram,
(b) radius-temperature diagram. Data from the radius expansion (0s--6s),
envelope contraction (6s--19s) and blackbody cooling curve (19s--100s)
are plotted with open triangles, filled triangles, and crosses
respectively.  Note the almost-horizontal track in the upper diagram
corresponding to the probable Eddington luminosity of the source, and
the close correspondence between the expansion and contraction
profiles in each diagram.
}
\end{center}
\end{figure}

In Figure~4a we present the luminosity-temperature ($L_{bol}$-$kT_c$) 
diagram for the 2000 burst, constructed using the parameters derived from
the fits to the spectra out to 100s after the burst rise. The
radius expansion and contraction phases are
almost horizontal at a luminosity corresponding to the
Eddington limit, with a slope introduced by general relativistic
effects.  Figure~4a can be directly compared to the equivalent
diagram for the 1988 burst (van Paradijs et al.\ 1990, Figure~9); in
removing the time domain, this representation emphasises the
similarity between the two bursts.  Figure~4b shows the strong inverse
correlation between the apparent blackbody radius and temperature; the track
defined by the envelope contraction lies almost on top of the
expansion track, indicating the reversibility of the thermodynamic
processes involved. From first principles the luminosity, radius, and
temperature of the photosphere should be related by the expression
$L=4\pi \sigma R^2 T^4$. In this case the luminosity is not quite
constant, and the radius and temperature obey the relation $R \propto
T^{-\alpha}$, with $\alpha$=2.06$\pm$0.04 (90\% confidence). Both
diagrams show the radiative cooling curve of the neutron star surface
after the touchdown of the photosphere.

While a simple blackbody often provides reasonable fits to burst
spectra from neutron stars, the energy distribution of the 'true'
emission is likely to be more complex.  A systematic difference is
expected between the blackbody color temperature $kT_c$ (=$kT_{bb}$)
obtained from spectral fitting and the effective temperature
$kT_{eff}$, due to the dominant contribution of electron scattering in
the hot neutron star atmosphere, and the resulting suppression of
emissivity (van Paradijs 1982). Model atmosphere calculations
show that the spectral hardening ratio $t$=$T_c/T_{eff}$ is greater
than 1, and increases as a function of $T_{eff}$ and of
$l$=$L_x/L_{Edd}$ (London et al.\ 1984, 1986; Ebisuzaki \& Nomoto
1986, 1988; Titarchuk 1994; Madej, 1997). The neutron star radius
observed by a distant observer, $R_\infty$, is related to $R_{bb}$ by
$R_{\infty}$=$R_{bb}t^2$.  Accurate knowledge of the behavior of $t$
would place direct constraints on the mass-radius relationship for
neutron stars.  However, there is as yet no convincing agreement
between the various existing theoretical models, and in addition the
model atmosphere calculations do not match the observed variations
with luminosity in several sources, including X1608$-$522 (Penninx et
al. 1989) and \source\ itself (see below, and Figure~5). If we adopt a
representative spectral hardening factor of $t$=1.5$\pm$0.1
(e.g. London et al.\ 1984, 1986; Madej 1997) for illustrative
purposes, our measured photospheric touchdown radius of
$R_{bb}$=8.6$\pm$1.0~km would imply $R_{\infty}$=19.3$\pm$3.4~km.

The value of the Eddington luminosity measured by a distant observer
depends upon the mass and radius of the emitting object. In the
limiting case, the Eddington luminosity measured at very large
photospheric radii, $L_{Edd, \infty} (R \gg R_\ast)$, will be greater
than the value measured at the point of photospheric touchdown,
$L_{Edd, \infty} (R = R_\ast)$, by a factor equivalent to the
gravitational redshift at the neutron star surface, $1+z_\ast$
(Paczy\'nski \& Anderson 1986; Damen et al.\ 1990).  At intermediate
radii, the ratio of the Eddington luminosities observed 
($L_{Edd, \infty} (R > R_\ast)/L_{Edd, \infty} (R = R_\ast))$ 
is a function of both the redshift and photospheric radius, as
modified by the (possibly varying) spectral hardening (Damen et al.\
1990). Therefore, the photospheric expansion track allows us to directly
estimate the gravitational redshift factor at the surface of the
neutron star, if we account correctly for the finite size of the
photosphere and for the temperature dependence of the scattering
opacity mentioned above.

For each data point in the photospheric expansion track (Figure~4a),
we applied Eqns.\ 7--9 of Damen et al.\ (1990) to obtain an
instantaneous estimate of ($1+z_\ast$), and then took the mean of
these derived values. For this calculation we adopted an Eddington
luminosity at touchdown of 
$L_{Edd, \infty} (R = R_\ast)$=(5.08$\pm$0.31)$\times$10$^{38}$
erg~s$^{-1}$ and a temperature at touchdown of 2.70$\pm$0.13~keV.

\begin{figure}[htb]
\figurenum{5}
\begin{center}
\begin{tabular}{c}
\psfig{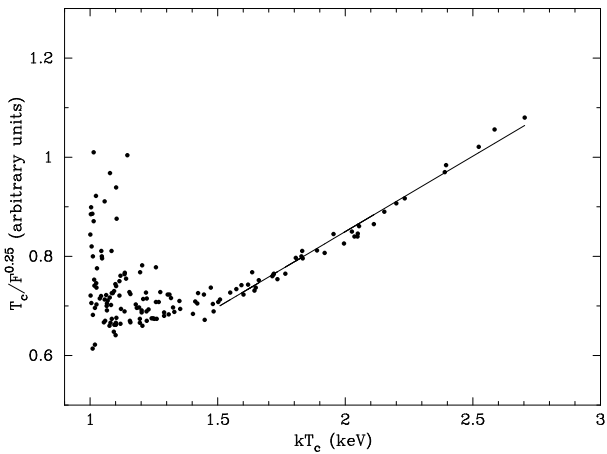}
\end{tabular}
\caption{The variation of the spectral hardening factor
$t$=$T_c/T_{eff}$ $\propto$ $T_c/F^{0.25}$, as a function of the color
temperature $T_{c}$ derived from spectral fits. The best-fitting
straight line for the higher-temperature data points is overlayed.
}
\end{center}
\end{figure}

Model atmosphere calculations during the expansion phase of a burst
have not yet been performed, so we calculated the redshift for
two differing assumptions. First, we assumed that the spectral
hardening ratio was constant during the expansion, and thus that no
correction was required for non-Planckian shape of the burst
spectra. For this case, we obtained a value of
$1+z_\ast$=1.17$\pm$0.05 (s.d.).
In the second derivation we assumed (again following Damen et al.\
1990) that the dependence of $t$ on the observed color temperature
$T_{c, \infty}$ is the same during radius expansion and contraction as
it is during the neutron star cooling. We cannot derive {\sl absolute}
values for $t$ from observational data alone; we can, however,
determine {\sl relative} values, and this is all the method requires.
During cooling, the effective temperature $T_{eff, \infty}$ is
proportional to $F_{\infty}^{0.25}$, and thus the hardening factor $t$
is proportional to $T_{c, \infty}/F_{\infty}^{0.25}$ (Penninx et al.\
1989; Damen et al.\ 1990). In Figure~5 we show $T_{c,
\infty}/F_{\infty}^{0.25}$ plotted against $T_{c, \infty}$ for the
cooling track of our \source\ burst, following photospheric touchdown.
For temperatures $kT_c>1.5$~keV, the relationship can be approximated
with a straight-line fit. Using this fit to scale the radius estimates
that enter the redshift calculation (Damen et al.\ 1990, Eqn.\ 9), our
derived gravitational redshift factor becomes $1+z_\ast$=1.28$\pm$0.06
(s.d.).

Applying the relativistic correction and solving for $L_{Edd}$ and
mass (e.g. Lewin et al.\ 1983, van Paradijs et al.\ 1990), we then
obtain, for the constant-$t$ case, a corrected Eddington luminosity of
$L_{Edd, \infty} (R \gg R_\ast)$=(5.94$\pm$0.45)$\times$10$^{38}$
erg~s$^{-1}$ and a neutron star mass of $M$=2.38$\pm$0.18$(1+X)
M_\odot$, where $X$ is the hydrogen content by mass; $X$=0 for
hydrogen-poor material and $X$=0.73 for cosmic composition). For
the variable-$t$ instance, these values are $L_{Edd, \infty} (R \gg
R_\ast)$=(6.50$\pm$0.50)$\times$10$^{38}$ erg~s$^{-1}$ and
$M$=2.60$\pm$0.22$(1+X) M_\odot$.

We emphasise that although only statistical errors are quoted here,
all current derivations of basic neutron star parameters from burst
spectroscopy are affected by the systematic uncertainties in the
interpretation of the spectral fit temperatures (van Paradijs et al.\
1990, Lewin et al.\ 1993). Our own data (Figure~5) imply a possible
variation in $t$ of $\sim$50\% over the temperature range 1.5$>kT_c>$2.7~keV
which is not well reproduced in the model atmosphere calculations.
The composition of the accreted matter, and degree of anisotropy in
the burst radiation, are additional key unknowns limiting our ability
to obtain the mass-radius relationship of the neutron star using these
methods.

Nonetheless, while higher than canonical, the values of $L_{Edd}$ and
$M$ derived here are consistent with those obtained by van Paradijs et
al.\ (1990) for the 1988 burst.  It therefore seems likely that the only
significant difference between the two bursts is in the overall
timescale of the burst, and thus the total amount of matter ignited in
the thermonuclear flash.  The quantity of energy liberated in the
combustion of hydrogen and helium into iron-peak elements depends on
the ratio of the original mix. Following Lewin et al.\ (1993) and
adopting a conversion fraction of 6$\times$10$^{18}$~erg~g$^{-1}$ =
0.007$c^2$, we calculate that the two events respectively consumed
1.1$\times$10$^{22}$g and 2.8$\times$10$^{21}$g of accumulated
material, each equivalent to many days of accretion from the
secondary.

\acknowledgments

We thank Cynthia Hess, Erik Kuulkers, and Tod Strohmayer for useful
comments and discussions.


\begin{references}

\reference{} Auri\`ere, M., Le F\`evre, O., \& Terzan, A., 1984, A\&A,
138, 415

\reference{} Bradt, H.V., Rothschild, R.E., \& Swank, J.H., 1993,
A\&AS, 97, 355

\reference{} Damen, E., Magnier, E., Lewin, W.H.G., Tan, J., Penninx,
W., \& van Paradijs, J., 1990, A\&A, 237, 103

\reference{} Day, C.S.R., \& Done, C., 1991, MNRAS, 253, 35P

\reference{} Durrel, P.R., \& Harris, W.E., 1993, AJ, 105, 1420

\reference{} Dotani, T., Inoue, H., Murakami, T., Nagase, F., Tanaka,
Y., Tsuru, T., Makishima, K., Ohashi, T., \& Corbet, R.H.D., 1990,
Nature, 347, 534

\reference{} Ebisuzaki, T. \& Nomoto, N., 1986, ApJ, 305, L67

\reference{} Ebisuzaki, T. \& Nomoto, N., 1988, ApJ, 328, 251

\reference{} Fabian, A.C., Pringle, J.E., \& Rees, M., 1975, MNRAS,
172, 15P

\reference{} Geisler, D., Minniti, D., Claria, J.J., 1992, AJ, 104, 627

\reference{} Harris, W.E., 1996, AJ, 112, 1487

\reference{} Hoffman, J.A., Cominsky, L., \& Lewin, W.H.G., 1980, ApJ,
240, L27

\reference{} Homer, L., \& Charles, P.A., 1998, New Astron. 3/7, 435

\reference{} Hut, P., McMillan, S., Goodman, J., Mateo, M., Phinney,
E.S., Pryor, C., Richer, Harvey B., Verbunt, F., \& Weinberg, M.,
1992, PASP, 104, 981

\reference{} Ilovaisky, S.A., Auri\`ere, M., Koch-Miramond, L.,
Chevalier, C., Cordoni, J.-P., \& Crowe, R.A., 1993, A\&A, 270, 139

\reference{} Jahoda, K., Swank, J. H., Giles, A. B., Stark, M. J.,
Strohmayer, T., Zhang, W., \& Morgan, E. H., 1996, in EUV, X-ray and
Gamma-Ray Instrumentation for Astronomy VII, ed O. H. Siegmund
(Bellingham, WA: SPIE), 59

\reference{} Lewin, W.H.G., van Paradijs, J., \& Taam, R.E., 1993,
SSR, 62, 223

\reference{} London, R.A., Taam, R.E., \& Howard, W.M., 1984, ApJ, 287, L27

\reference{} London, R.A., Taam, R.E., \& Howard, W.M., 1986, ApJ, 306, 170

\reference{} Madej, J., 1997, A\&A, 320, 177

\reference{} Nakamura, N., Dotani, T., Inoue, H., Mitsuda, K., Tanaka,
Y., \& Matsuoka, M., 1989, PASJ, 41, 617

\reference{} Nishimura, J., Mitsuda, K., \& Itoh, M. 1986, PASJ, 38,
819

\reference{} Paczy\'nski, B., \& Anderson, N., 1986, ApJ, 302, 1

\reference{} Penninx, W., Damen, E., Tan, J., Lewin, W.H.G., \& van
Paradijs, J., 1989, A\&A, 208, 146

\reference{} Sidoli, L., Parmar, A.N., \& Oosterbroek, T., 2000, A\&A,
360, 520

\reference{} Sneden, C., Kraft, R.P., Prosser, C.F., \& Langer, G.E.,
1991, AJ, 102, 2001

\reference{} Titarchuk, L., 1994, ApJ, 429, 340

\reference{} van Paradijs, J., 1982, A\&A, 107, 51

\reference{} van Paradijs, J., Dotani, T., Tanaka, Y., \& Tsuru, T.,
1990, PASJ, 42, 633

\reference{} van Paradijs, J., Penninx, W., \& Lewin, W. H. G. 1988,
MNRAS, 233, 437

\end{references}
\end{document}